\documentclass[aps,nofootinbib,preprint]{revtex4}
\usepackage{amsbsy}
\usepackage{epsf}

\usepackage{amsmath}
\usepackage{graphicx}
\usepackage{amssymb}
\usepackage{color}
\usepackage{soul}
\def\journalfont{\rm}         
\def\jou#1{{\journalfont #1\ }}
\def\joudef#1#2{\def #1{\jou{\ignorespaces #2}}}

\joudef{\aaa}    { Astron.\ Astrophys.}
\joudef{\aip}    { Adv.\ Phys.}
\joudef{\adm}    { adv.\ math.}
\joudef{\am}     { Ann.\ Math.}
\joudef{\apl}    { Ann.\ Phys.\ (Leipzig)}
\joudef{\apny}   { Ann.\ Phys.\ (N.Y.)}
\joudef{\arnps}  { Annu.\ Rev.\ Nucl.\ Part.\ Sci.}
\joudef{\apj}    { Astrophys.\ J.}
\joudef{\apjl}    { Astrophys.\ J.\ Lett.}
\joudef{\cjp}    { Can.\ J.\ Phys.}
\joudef{\cmp}    { Commun.\ Math.\ Phys.}
\joudef{\cqg}    { Class.\ Quantum Grav.}
\joudef{\grg}    { Gen.\ Rel.\ Grav.}
\joudef{\ijmpd}  { Int.\ J.\ Mod.\ Phys.\ D}
\joudef{\ijtp}   { Int.\ J.\ Theor.\ Phys.}
\joudef{\invm}   { Invent.\ Math.}
\joudef{\jm}     { J.\ Math.}
\joudef{\jmaa}   { J.\ Math.\ Anal.\ Appl.}
\joudef{\jmp}    { J.\ Math.\ Phys.}
\joudef{\jpa}    { J.\ Phys.\ A}
\joudef{\lr}    { Liv.\ Rev.\ Rel.}
\joudef{\mnras}  { Mon.\ Not.\ R.\ Ast.\ Soc.}
\joudef{\mpl}   { Mod.\ Phys.\ Lett.} 
\joudef{\mpla}   { Mod.\ Phys.\ Lett.\ A} 
\joudef{\nature} { Nature}
\joudef{\nc}     { Nuovo Cim.}
\joudef{\npb}    { Nuc.\ Phys.\ B}
\joudef{\ph}     { Physica}
\joudef{\pla}    { Phys.\ Lett. A}
\joudef{\plb}    { Phys.\ Lett. B}
\joudef{\pr}     { Phys.\ Rev.}
\joudef{\pra}    { Phys.\ Rev.\ A}
\joudef{\prb}    { Phys.\ Rev.\ B}
\joudef{\prc}    { Phys.\ Rev.\ C}
\joudef{\prd}    { Phys.\ Rev.\ D}
\joudef{\prep}   { Phys.\ Rep.}
\joudef{\prl}    { Phys.\ Rev.\ Lett.}
\joudef{\prsla}  { Proc.\ Roy.\ Soc.\ Lond.\ A}
\joudef{\ptp}    { Prog.\ Theor.\ Phys.}
\joudef{\ptps}   { Prog.\ Theor.\ Phys.\ Suppl.}
\joudef\rmp      { Rev.\ Mod.\ Phys.}
\joudef\spj      { Sov.\ Phys.\ JETP}

\newcommand{\ncd}{\newcommand}
\ncd{\beq} {\begin{equation}}
\ncd{\eeq} {\end{equation}}

\def\x{{\mathrm{x}}}
\def\y{{\mathrm{y}}}

\def\n{{\rm f}}

\def\N{\mathrm{c}}
\def\({\left(}
\def\){\right)}
\def\[{\left[}
\def\]{\right]}

\ncd{\dell}{\partial}
\ncd{\nfrac}[2]{\left(\frac{n_{#1}}{n_{#2}}\right)^2}
\ncd{\pc}{\check{p}}
\ncd{\rhoc}{\check\rho}
\ncd{\betac}{\check\beta}
\ncd{\muc}{\check\mu}
\ncd{\Oc}{\check\Omega}
\ncd{\ec}{\check\varepsilon}
\ncd{\tsfrac}[2]{{\textstyle\frac{#1}{#2}}}

\def\csc{\check{c}^2_s}

\newcommand{\bear}{\begin{eqnarray}}
\newcommand{\eear}{\end{eqnarray}}
\newcommand{\A}{\mathcal{A}}
\newcommand{\B}{\mathcal{B}}

\ncd{\bldeta}{\boldsymbol{\eta}}
\ncd{\bldone}{\mathbf{1}}
\ncd{\blds}{\mathbf{s}}
\ncd{\bldk}{\mathbf{k}}
\ncd{\blde}{\mathbf{e}}

\ncd{\abs}[1] {|#1|}
\ncd{\ubold}{\mathbf u}
\ncd{\Abold}{\mathbf A}
\ncd{\Bbold}{\mathbf B}
\ncd{\Mbold}{\mathbf M}
\ncd{\lagom}{\hspace{.6pt}}
\ncd{\muk}{k}
\ncd{\lagomdot}{{\mbox{\large$\cdot$}}}

\ncd{\stil}{\tilde{s}}
\ncd{\ftil}{\tilde{f}}
\ncd{\Otil}{\tilde{\Omega}}

\ncd{\D}{\mathcal{D}}
\ncd{\Qcal}{\mathcal{Q}}
\ncd{\Jcal}{\mathcal{J}}
\ncd{\Ecal}{\mathcal{E}}
\ncd{\Wcal}{\mathcal{W}}
\ncd{\Xcal}{\mathcal{X}}
\ncd{\Ycal}{\mathcal{Y}}
\ncd{\Bcal}{\mathcal{B}}
\ncd{\Acal}{\mathcal{A}}
\ncd{\Scal}{\mathcal{S}}
\ncd{\Ccal}{\mathcal{C}}

\ncd{\vt}{v_{t\perp t}^{\,2}}

\ncd{\psimap}{\Psi}
\ncd{\psivar}{\psi}

\ncd{\Ap}{(r\psivar)'}
\ncd{\Adot}{(r\psivar)^{\lagomdot}}
\ncd{\Bp}{(r^{-1}\varphi)'}
\ncd{\Bdot}{(r^{-1}\varphi)^{\lagomdot}}

\ncd{\ela}{\left(1-\frac{2m}{r}\right)}

\ncd{\shm}{S}
\ncd{\shmtwoD}{\mathcal{\shm}}
\ncd{\lie}{\mathcal{L}}
\ncd{\brk}{\mathrm{max}}
\ncd{\fgauge}{f_\mathrm{G}}
\ncd{\I}{\mathrm{c}}
\ncd{\f}{\mathrm{f}}

\begin{document}

\title{The dynamics of neutron star crusts: Lagrangian perturbation theory for a relativistic superfluid-elastic system}

\author{N.~Andersson$^1$, B.~Haskell$^2$,  G.L.~Comer$^3$ and L. Samuelsson$^4$}

\affiliation{$^1$ Mathematical Sciences and STAG Research Centre,
University of Southampton, Southampton SO17 1BJ, United Kingdom \\
$^2$ Nicolaus Copernicus Astronomical Center, Polish Academy of Sciences, ul. Bartycka 18, 
00-716 Warsaw, Poland \\
$^3$ Department of Physics, Saint Louis University, St.~Louis, MO, 63156-0907, USA \\
$^4$ Nordita, Roslagstullsbacken 23, SE-106 91 Stockholm, Sweden}

\begin{abstract}
The inner crust of a mature neutron star is composed of an elastic lattice of neutron-rich nuclei 
penetrated by free neutrons. These neutrons can flow relative to the crust once the star cools 
below the superfluid transition temperature. In order to model the dynamics of this system, which 
is relevant for a range of problems from pulsar glitches to magnetar seismology and continuous 
gravitational-wave emission from rotating deformed neutron stars, we need to understand 
general relativistic Lagrangian perturbation theory for elastic matter coupled to a superfluid 
component.  This paper develops the relevant formalism to the level required for astrophysical 
applications.
\end{abstract}

\maketitle

\section{Introduction}

Astrophysical observations of neutron stars provide an important probe of the state of matter 
under extreme  conditions. Shortly after the star is born, the outer layers freeze to form an 
elastic crust and the temperature of the high-density core drops below the level where superfluid 
and superconducting components are expected to be present. The different phases of matter 
impact on the observed phenomenology in a variety of ways. The crust region is important as it 
anchors the star's magnetic field (and provides specific channels for the gradual field evolution 
\cite{2013MNRAS.434..123V}), leading to an immediate connection between observed quasi-
periodic oscillations in the tails of magnetar flares \cite{Watts06:_xray_oscs} and the dynamics of 
the elastic nuclear lattice. A detailed understanding of the properties of the crust is essential for 
 efforts to match the theory to observed seismology features 
\cite{Samuelsson07:_axial_crust,2009CQGra..26o5016S}. In a different context, the ability of the crust to sustain elastic 
strain is key to the formation of asymmetries which may lead to detectable gravitational waves 
from a mature spinning neutron star. Continuous gravitational-wave searches with the LIGO-Virgo network of interferometers is beginning to set interesting upper limits for such signals for a 
number of known pulsars \cite{2017ApJ...839...12A}, in some instances reaching significantly 
below the expected maximum ``mountain'' size estimated from state of the art molecular 
dynamics simulations of the crustal breaking strain 
\cite{2009PhRvL.102s1102H,2013PhRvD..88d4004J}. Finally, the ability of the neutron-rich 
nuclei in the crust to pin superfluid vortices is also a key part of the standard explanation for 
observed glitches in young pulsars \cite{2015IJMPD..2430008H,2016MNRAS.455.3952S}. All things considered, the 
crust region is crucial for an understanding of neutron star phenomenology and we need to make 
sure that our theoretical models incorporate as much of the relevant physics as possible. 

As an important step towards the development of an appropriate theoretical 
framework, we will extend Lagrangian perturbation theory to the coupled superfluid-elastic 
crust system in the context of general relativity (extending the framework used in 
\cite{1978CMaPh..62..247F} to discuss the gravitational-wave driven instability of rotating 
relativistic stars, see \cite{na03:_rev} for a review). This is a formal development, but it should be 
immediately relevant to efforts aimed at modelling specific astrophysical scenarios. It is natural to 
use a Lagrangian framework since the perturbations of the elastic component becomes much 
``simpler'' when considered in a frame that is co-moving with the crust. Moreover, it is essential 
that the problem is considered in the framework of general relativity as this is a pre-requisite for 
any quantitative analysis based on a realistic matter description. However, as the  combined 
superfluid-elastic problem is still rather complex (and it is helpful to make the development as 
clear and intuitive as possible), we will construct the theory step by step, starting with a review of a single perfect fluid, considering next a pure elastic crust, and then adding 
the anticipated superfluid neutron component (as well as the associated entrainment effect 
\cite{Carter04:_crust}).

We take as our starting point the series of papers by Karlovini and Samuelsson 
\cite{Karlovini03:_elas_ns_1,Karlovini04:_elas_ns_2,Karlovini04:_elas_ns_3,Karlovini07:_elas_ns_4}, 
which build on earlier work by Carter and Quintana \cite{Carter72:_gr_elas} (see also 
\cite{carter06:_crust,carter73:_elast_pertur,Carter04:_bcs,carter06:_ent_nscrust}), and the 
convective variation approach to relativistic fluid dynamics 
\cite{carter89:_covar_theor_conduc,andersson07:_livrev}. The main focus of our discussion is 
the intimate connection between relativistic elasticity and Lagrangian perturbation theory.  This 
link has not previously been explored in detail, yet we will demonstrate that  Lagrangian 
variations capture neutron star crust physics in a natural fashion. 

Throughout the discussion we assume a spacetime represented by a metric $g_{ab}$ with  
signature  $(-,+,+,+)$ and use early Latin letters, $a, b, c, d \ldots$ to denote abstract spacetime 
indices. The Einstein summation convention applies, unless otherwise stated.

\section{The variational approach}

Our main focus may be on elastic matter, but it is nevertheless  natural to begin by reviewing the 
variational approach to relativistic fluid dynamics
\cite{carter89:_covar_theor_conduc,andersson07:_livrev}. This is useful for two reasons: First of 
all, it is important to understand this formulation in order to extend it beyond simple perfect fluid 
models, e.g. add elasticity and additional fluid components that should be present when the 
system becomes superfluid. Secondly, the variational derivation already involves Lagrangian 
variations. Hence, the derivation of the fluid equations of motion provides some of the results we 
need if we want to study Lagrangian perturbations of a more general system. 

In order to avoid undue confusion, let us consider the simplest model: a single barotropic fluid. In 
this case the matter equation of state can be expressed in terms of an energy functional that is a 
function of a single parameter. We  take this parameter to be the particle number density, $n$, 
and assume that the dynamics is governed by a Lagrangian $\Lambda (n)$. The relation 
between this Lagrangian and the energy of the system will become clear shortly. The matter flux 
is represented by a (conserved) flux $n^a$, such that $n^2 = -g_{ab} n^a n^b$. In effect, this 
means that the Lagrangian depends on the flux and the spacetime metric. 

An arbitrary variation of $\Lambda=\Lambda(n^2)=\Lambda(n^a,g_{ab})$ then gives (ignoring 
terms that can be written as total derivatives, that is,  ``surface terms'', in the action \cite{andersson07:_livrev})
\beq
    \delta \left(\sqrt{- g} \Lambda\right) = \sqrt{- g} \left[\
    \mu_a \delta n^a + \frac{1}{2} \left(\Lambda g^{a b} +  
    n^a \mu^b\right) \delta g_{a b}\right] \ , 
    \label{dlamb} 
\eeq
where $g$ is the determinant of the spacetime metric and $\mu_a$ is the canonical momentum 
defined by
\beq
     \mu_{a} = {\partial \Lambda \over \partial n^a} = -2 {\partial \Lambda \over \partial n^2} g_{ab} n^b = \mathcal B n_a \ .
     \label{momdef}
                   \end{equation} 
We  have  also used
\begin{equation}
\delta \sqrt{-g} =  {1\over 2} \sqrt{-g} g^{ab} \delta g_{ab}  \  .
\end{equation}

The result in \eqref{dlamb} shows why a variational derivation of fluid dynamics is nontrivial. As it 
stands, the variation of $\Lambda$ suggests that the equations of motion should be $\mu_a=0$. 
In essence,  the fluids would not carry energy or momentum. This is obviously not what we are 
looking for. To resolve this issue, we need a constrained variation. We need to insist that the matter flux is 
conserved. That is, we want to ensure that 
\beq
\nabla_a n^a = 0 \ .
\eeq
A natural way to do this is to make use of a three-dimensional ``matter space'' \cite{carter89:_covar_theor_conduc}. The coordinates 
of  this matter space, $X^A$ where $A = \{1,2,3\}$, serve  as  labels that distinguish individual  
fluid element worldlines \cite{andersson07:_livrev}. These labels are assigned at the initial time of 
the evolution, say $t=0$. The matter space coordinates can be considered as scalar fields on 
spacetime, with a unique map (obtained by a pull-back construction) relating them to the 
spacetime coordinates. 

The variational construction then involves three  steps. First we note that the conservation of the 
individual fluxes is ensured provided the dual three-form 
\begin{equation}
 n_{a b d} = \epsilon_{a b d e} n^e   \quad , \quad
    n^a = \frac{1}{3!} \epsilon^{a b d e} n_{b d e} \ , 
    \label{n3form}
 \end{equation}
(where $\epsilon_{a b d e}$ is the usual volume form associated with the 
spacetime) is closed, i.e.
\begin{equation} 
     \nabla_{[a} n_{b d e]} = 0 \quad \longrightarrow \quad 
  \nabla_{a} n^{a} = 0 \ .
   \label{consv2} 
\end{equation} 
In the second step we make use of the matter space to construct three-forms that are 
automatically closed on spacetime, i.e.
\beq
    n_{a b d} =  \psi^A_{[a}\psi^B_b\psi^D_{d]}
                   n_{A B D} \ , \label{pb3form}
\eeq 
(the square brackets indicate anti-symmetrization, as usual) where the map is given by
\beq
\psi^A_a = {\partial X^A\over \partial x^a} \ ,
\eeq
and the Einstein summation convention applies to repeated matter-space indices $\{A,B,...\}$. 
The volume form $n_{A B D}$, which is anti-symmetric, provides matter space with a geometric 
structure (we elaborate on this in the Appendix). If integrated over a volume in matter space it 
provides a measure of the number of particles in that volume. With this definition, the three form 
\eqref{pb3form} is closed if $n_{A B D}$ is a function only of the $X^A$. In other words, the 
scalar fields (in spacetime) $X^A$  are taken to be fundamental variables. A consequence of 
this construction is that $n_{A B D}$ is ``fixed'' on matter space\footnote{Each point in matter 
space is associated with a particular worldline in spacetime. The matter space coordinates are a 
set of three scalar fields on spacetime, such that their values do not change along their particular 
worldline. As the fluid evolves in spacetime, any object in matter space which depends on only its 
own matter space coordinates (i.e.~a tensor) will therefore not change its value at each 
coordinate. In this sense, it is ``fixed'', even though the object itself can vary across matter space 
points. As a side note, Andersson and Comer \cite{Andersson15:_dissfl_act} have shown that 
when $n_{A B D}$ is no longer fixed, an action principle incorporating dissipation can be built.}.

The final step involves introducing the  Lagrangian displacement, $\xi^a$, and linking back to the 
spacetime perturbations. The displacement tracks the movement of a given fluid element. From 
the standard definition of Lagrangian variations in the relativistic context,  we  have 
\beq
\Delta X^A = \delta X^A + \mathcal{L}_{\xi} X^A = 0 \ , 
\label{DelX}
\eeq
where $\delta X^A$ is the Eulerian variation and $\mathcal{L}_{\xi}$ is the Lie derivative along 
$\xi^a$. We see that, convective variations are such that (since $ X^A$ is a scalar field on 
spacetime)
\beq
  \delta X^A = -  \mathcal{L}_{\xi} X^A = - \xi^a {\partial X^A \over \partial x^a} = -\xi^a \psi^A_a\ . 
    \label{xlagfl}
\end{equation} 
For later benefit, it is worth noting that this leads to
\beq
\Delta \psi^A_a = 0 \ .
\eeq
After some algebra, one finds  
\beq
\Delta n_{a b d} = 0 \ , 
\label{pertform}
\eeq
which in turn implies
\beq
\delta n^a  = n^b \nabla_b \xi^a - \xi^b 
                   \nabla_b n^a - n^a \left(\nabla_b 
                   \xi^b + \frac{1}{2} g^{b d} \delta 
                   g_{b d}\right) = - \mathcal{L}_{\xi} n^a - n^a \left(\nabla_b 
                   \xi^b + \frac{1}{2} g^{b d} \delta 
                   g_{b d}\right)  \ . 
                   \label{delnvec} 
\end{equation} 
This is the main result of the exercise.

Now we can return to the variation of the matter Lagrangian. By expressing the variation of 
$\Lambda$ in terms of the displacement $\xi^a$ we ensure that the flux conservation is 
accounted for in the equations of motion. We get
\beq\
    \delta \left(\sqrt{- g} \Lambda\right) = \sqrt{- g} \left\{ \frac{1}{2}\left[\left( \Lambda - n^d \mu_d\right) g^{a b} +  
    n^a \mu^b \right] \delta g_{a b} + f_a \xi^a \right\} \ , \label{variable}
\eeq
and it  follows that the equations of motion are given by
\begin{equation} 
     f_b \equiv 2 n^a \nabla_{[a} \mu_{b]} = 0  \ .
          \label{force}
\eeq
Meanwhile, the stress-energy tensor follows as
\begin{equation}
T^{ab} = {2 \over \sqrt{-g}} {\delta \left( \sqrt{-g}\Lambda\right) \over \delta g_{ab}} =  \left( \Lambda - n^d \mu_d\right) g^{a b} +  
   n^a\mu^b \ .
\label{stressen1}
\end{equation}

The final results may seem somewhat unfamiliar, but it is easy to recast them in a more 
commonly used form. All we need is a bit of thermodynamics. First we introduce the matter 
four-velocity such that $n^a=nu^a$. Then it follows that the chemical potential is, c.f. 
Eq.~\eqref{momdef},
\beq
- u^a \mu_a =  \mu = n \Bcal \ ,
\eeq
Moreover, an observer moving with the matter flow would measure the mass-energy
\begin{equation}
\varepsilon = u_a u_b T^{ab} = - \Lambda \ ,
\end{equation}
which means that 
\beq
\mu = {d\varepsilon \over dn} \ ,
\eeq
as expected.

The fundamental relation \cite{reichl98:_book}
\begin{equation}
p = - \varepsilon + n\mu  =  \Lambda - n^a \mu_a \ ,
\label{fundrel}
\end{equation}
which defines the pressure,  means that we have
\begin{equation}
T^{ab} = p g^{ab} + n^a \mu^b = \varepsilon u^a u^b + p h^{ab}\ ,
\label{stressen1}
\end{equation}
where we have introduced the standard spacetime projection
\begin{equation}
h^{ab} = g^{ab} + u^a u^b \ .
\end{equation}
Not surprisingly, Eq.~\eqref{stressen1} is the usual  perfect fluid stress-energy tensor.

Next, let us consider the equations of motion \eqref{force}. Making use of our various definitions, 
the momentum equation can be written
\beq
\mu \dot{u}_b + h^a_b \nabla_a \mu = 0 \ ,
\eeq
where $\dot{u}_b=u^a \nabla_a u_b$ is the four acceleration. Again making use 
Eq.~\eqref{fundrel}, we arrive at the standard relativistic Euler equation. That is,
\beq
\dot{u}_b + { 1 \over p+\varepsilon} h^a_b \nabla_a p = 0 \ .
\eeq

An easy way to see that this result was inevitable is to note that 
\begin{equation}
\nabla_a T^{ab} = f^b + \nabla^b \Lambda - \mu_a \nabla^b n^a = f^b = 0 \ .
\end{equation}
The second equality follows from i) the fact that $\Lambda$ is a function only of $n^a$ and 
$g_{ab}$, and ii) the definition of the momentum $\mu_a$.

\section{Lagrangian perturbations} \label{langperb}

By introducing the displacement $\xi^a$, effectively tracking the fluid elements, we have 
prepared the ground for a study of Lagrangian perturbations. In fact, we see immediately from 
\eqref{delnvec} that
\beq
\Delta n^a = - n^a \left( \nabla_b \xi^b + { 1 \over 2} g^{bd} \delta g_{bd} \right) = - {1 \over 2} n^a \left( g^{bd} \Delta g_{bd}\right)\ ,
\label{dna}
\eeq
where
\beq
\Delta g_{ab} = \delta g_{a b} + 2 \nabla_{(a} \xi_{b)} \ , \label{lagvarmet}
\eeq
(the parentheses indicate symmetrization).  Eq.~\eqref{dna} has a natural 
interpretation: The variation of a fluid worldline with respect to its own Lagrangian displacement 
has to be along the worldline and can only measure the changes of the volume of its own fluid 
element. This is one of the advantages of the Lagrangian variation approach,   alluded to earlier. 
It also follows that \cite{1978CMaPh..62..247F}
\beq
\Delta n = -{ n \over 2} h^{ab} \Delta g_{ab} \ ,
\label{dn}\eeq
and
\beq
\Delta u^a = { 1 \over 2} u^a u^b u^d \Delta g_{bd} \ .
\eeq

For any given equation of state $\Lambda(n)$, we can now express the perturbed equations of 
motion in terms of the displacement vector $\xi^a$ and the Eulerian variation of the metric 
$\delta g_{ab}$. In doing this it is worth noting that the usual approach to relativistic stellar 
perturbations is to work with this combination of variables (see for example 
\cite{1992PhRvD..46.4289K}). Essentially, we need the Eulerian perturbation of the Einstein field 
equations and the Lagrangian variation of the momentum equation \eqref{force}. The description 
of the perturbed Einstein equations is standard, so we focus on the fluid aspects here.

The perturbations of \eqref{force} are easy to work out once we note that the Lagrangian 
variation commutes with the exterior derivative.  We immediately get
\beq
(\Delta n^a) \nabla_{[a}\mu_{b]} + n^a \nabla_{[a}\Delta \mu_{b]} = 0 \ .
\label{pmom1}
\eeq
This simplifies further if we use \eqref{dna} and assume that the background is such that 
\eqref{force} is satisfied. The first term then vanishes, and we are left with
\beq
 n^a \nabla_{[a}\Delta \mu_{b]} = 0 \ .
\label{euler1}
\eeq
To complete this expression, we need to work out $\Delta \mu_a$. This is a straightforward task 
given the above results, and we find
\beq
\Delta \mu_a = \left( \mathcal{B} + n {d \mathcal{B} \over dn} \right) g_{ab} \Delta n^b + \left( \mu^b \delta_a^d - {d \mathcal{B} \over d n^2} n_a n^b n^d \right) \Delta g_{b d} \ .
\eeq
For later convenience, we note that this expression can be written \cite{andersson07:_livrev}
\beq
\Delta \mu _a = \mathcal{B}_{ab} \Delta n^b + {1 \over 2} g^{d b} \left(\delta^e_a \mu_b + \mathcal{B}_{a b} n^e \right) \Delta g_{d e} \ , \label{dmua1}
\eeq
where 
\beq
\mathcal{B}_{ab} = \mathcal{B} g_{ab} - 2 {d \mathcal{B} \over d n^2} n_a n_b \ .
\label{Bab}
\eeq
If we insert Eq.~\eqref{dna} into \eqref{dmua1}  we find
\begin{multline}
\Delta \mu_a= - {1 \over 2} \mathcal{B}_{ab} n^b \left(g^{d e} \Delta g_{d e}\right) + {1 \over 2} g^{d b} \left(\delta^e_a \mu_b + \mathcal{B}_{a b} n^e \right) \Delta g_{d e} \cr
= {1 \over 2} \left[\delta^e_a \mu^d + \mathcal{B}_{a b} \left(g^{d b} n^e - g^{d e} n^b\right)\right] \Delta g_{d e} \ . \label{dmua2}
\end{multline} 
That is,  $\Delta n^a$ has been completely replaced by $\Delta g_{a b}$ in the fluid equations, 
thus completing the point about the advantage of Lagrangian displacements. Finally, in order to 
interpret the perturbed momentum we  note that, in the single-fluid case the speed of sound 
follows from
\beq
c_s^2 = {dp \over d\varepsilon} = { n \over \mu} \left(  \mathcal{B} + n {d \mathcal{B} \over d n} \right) \ .
\eeq

In principle, we now have all the results we need in order to express the perturbed equations of 
motion \eqref{euler1} in terms of $\xi^a$ and $\delta g_{ab}$. In doing this, it is worth noting that 
\eqref{euler1} is orthogonal to $u^b$, which means that the problem only has three fluid degrees 
of freedom. This is natural since the conservation of particle number was guaranteed by the 
construction of the framework. 

Later, we will extend the analysis to  problems with several distinct fluid flows, as required to 
describe, for example, heat flux and/or systems with superfluid components 
\cite{andersson07:_livrev}. In addition, we want to account for the possibility that one of these 
components is elastic rather than fluid. At the end of the day, we want to arrive at a formulation 
that allows us to model the dynamics of a realistic neutron star crust. To reach this point, we 
need to extend the formalism in two directions: We need to i) account for the crust elasticity and 
ii) allow for the presence of a superfluid neutron component. For practical reasons, it makes 
sense to first consider the elasticity.

\section{Relativistic elasticity}
\label{relelas}

With some of the formalities out of the way, let us return to the variational derivation of the fluid 
equations, with the intention of extending the analysis to account for elasticity. The  motivation for 
this exercise is that ``force-balance equations'' like \eqref{force} are  readily (as we will see later) 
adapted to multi-fluid settings, where it is necessary to have individual momentum equations for 
the different constituents \cite{andersson07:_livrev}. Some of these equations can be extracted from the stress energy tensor, but this 
route is not as elegant and additional information would still be required.

The modern view of elasticity builds on the comparison of an actual matter configuration to an 
unstrained reference shape. In order to keep track of the reference state relative to which the 
strain is measured, we introduce a positive definite and symmetric tensor field $k_{a b}$ 
\cite{Karlovini03:_elas_ns_1}. Intuitively, this tensor encodes the (3-)geometry of the solid (as 
seen by the solid itself). The tensor $k_{a b}$ is similar to $n_{a b d}$ in the sense that it is flow-line orthogonal, $u^a k_{a b} = 0$, and fixed in matter space.  Moreover, as discussed in the 
Appendix, key properties of $k_{a b}$ are established by introducing the corresponding matter 
space object, $k_{A B} (= k_{B A})$, through
\beq
      k_{a b} = \psi^A_a \psi^B_b k_{A B} \ . \label{kAB}
\eeq
First of all, the Lagrangian variation of $k_{a b}$ vanishes [c.f.~Eq.~\eqref{Delkab}]. This means 
that $k_{a b}$, in addition to being a natural quantity for describing the elastic configuration, is 
 useful in the development of Lagrangian perturbation theory. In particular,
\beq
\mathcal L_u k_{ab} = 0 \ . \label{kabldrag}
\eeq
Next, by assuming that $n_{A B D}$ is the volume form associated with $k_{A B}$ 
[c.f.~Eq.~\eqref{kn3form}], one can show that $k$ (the determinant of $k_{a b}$) is such 
that $k = n^2$ \cite{Karlovini03:_elas_ns_1}, even though $k_{a b}$ does not depend on the 
number density $n$. 

Letting the Lagrangian $\Lambda$ depend also on this new tensor (in essence, incorporating the 
energy associated with elastic strain) we have
\beq
    \delta \left(\sqrt{- g} \Lambda\right) = \sqrt{- g} \left[
    \mu_a \delta n^a + \left( \frac{1}{2}\Lambda g^{a b} +  
    {\partial \Lambda \over \partial g_{ab}} \right) \delta g_{a b} + {\partial \Lambda \over \partial k_{ab} }\delta k_{ab} \right] \ . \label{dlamb2}
\eeq
We proceed as before and replace $\delta n^a$ with the Lagrangian displacement $\xi^a$. In 
addition, we have from Eq.~\eqref{Delkab} in the Appendix
\begin{equation}
       \delta k_{ab} = - \xi^d \nabla_d k_{ab} - k_{d b} \nabla_a \xi^d - k_{a d} \nabla_b \xi^d \ .
\end{equation}
Again ignoring surface terms, we have (as $k_{ab}$ is symmetric)
\begin{equation}
{\partial \Lambda \over \partial k_{ab} }\delta k_{ab} =  \xi^a \left[ 2 \nabla_b \left( {\partial \Lambda \over \partial k_{bd} } k_{a d}\right) - {\partial \Lambda \over \partial k_{bd} }\nabla_a k_{b d}  
\right] \ .
\end{equation}
Making use of this result, we arrive at
\beq
    \delta \left(\sqrt{- g} \Lambda\right) = \sqrt{- g} \left\{ \left[ \frac{1}{2}\left( \Lambda - n^d \mu_d\right) g^{a b} +  
    {\partial \Lambda \over \partial g_{ab}} \right] \delta g_{a b} + \tilde f_a \xi^a \right\} \ , \label{dlamb3}
\eeq
where
\begin{equation}
\tilde f_a = 2 n^b \nabla_{[a}\mu_{b]} + 2 \nabla_b \left( {\partial \Lambda \over \partial k_{bd} } k_{a d}
\right) - {\partial \Lambda \over \partial k_{bd} }\nabla_a k_{b d} = 0  \ .
\label{tforce}
\end{equation}
As in the fluid case, this result provides the equations of motion for the system. However, we  
need to do a bit of work in order to get the result into a user-friendly form. To start with, 
we read off the stress-energy tensor from \eqref{dlamb3}:
\begin{equation}
T^{ab} =  \left( \Lambda - n^d \mu_d\right) g^{a b} + 2 {\partial \Lambda \over \partial g_{ab}} \ .
\label{stressen}
\end{equation}

The next step involves  giving physical meaning to $k_{ab}$. As we want to model elasticity, we 
need to quantify the deviation of a given state from a relaxed configuration. In order to do 
this, it is convenient to follow Karlovini and Samuelsson \cite{Karlovini03:_elas_ns_1} and 
introduce one further matter space tensor, $\eta_{A B}$. This object depends on $n$, and relates 
directly to the relaxed state.  Its defining characteristic is that, in the relaxed configuration, it is 
the inverse to 
\beq
g^{AB} = \psi^A_a \psi^B_b g^{ab} =  \psi^A_a \psi^B_b h^{ab}
\eeq
That is, for this specific state, we have
\beq
      g^{A C} \eta_{C B} = \delta^A_B \ . \label{etainv1}
\eeq
The spacetime counterpart is
\beq
      \eta_{a b} = \psi^A_a \psi^B_b \eta_{A B} \ .
\eeq 
and, as outlined in the Appendix, one can show that \cite{Karlovini03:_elas_ns_1}  
\begin{equation}
\eta_{a b} = n^{- 2/3} k_{a b} \ . \label{etadef}
\end{equation}
This relation is important, as we have already established that $k_{ab}$ is a fixed matter space 
tensor and this will be crucial when we consider Lagrangian perturbations.

Let us now imagine that the system evolves away from the relaxed state. This means that 
\eqref{etainv1} no longer holds:  $\eta_{AB}$ retains the value set by the initial state, but 
$g^{AB}$ evolves along with the spacetime.  This leads to the build up of elastic strain, simply 
quantified in terms of the strain tensor
\begin{equation}
s_{a b} = {1\over 2} ( h_{a b} - \eta_{a b}) =  {1\over 2} \left( h_{a b} - n^{-2/3} k_{a b} \right)\ .
\label{sab}
\end{equation}
In the relaxed configuration, we have $\eta_{ab} = h_{ab}$ by construction so it is obvious that 
$s_{ab}$ vanishes. 

This description is quite intuitive, but in practice it is more natural to work with scalars formed 
from $\eta_{ab}$  (which can be viewed as ``invariant''). This makes the model less abstract. 
Hence we introduce  the  strain  scalar $s^2$ as a suitable combination (see below) of the 
invariants of $\eta_{ab}$:
\begin{eqnarray}
I_1 &=& \eta^a_{\ a} = g^{A B} \eta_{A B} 
         \ , \cr 
I_2 &=& \eta^a_{\ b} \eta^b_{\ a} = g^{A D} g^{B E} \eta_{E A} \eta_{D B} 
          \ , \cr 
I_3 &=& \eta^a_{\ b} \eta^b_{\ d} \eta^d_{\ a} = g^{A E} g^{B F} g^{D G} \eta_{E B} \eta_{F D} 
              \eta_{G A} 
\ . \label{invs}
\end{eqnarray}
However, because of the Cayley-Hamilton theorem \cite{birkhoff1965survey}, the number 
density $n$ also can be seen to be a combination of invariants, i.e.
\begin{equation}
     k = n^2 = {1\over 3!} \left( I_1^3 - 3 I_1I_2+2I_3 \right) \ . 
\end{equation}
Thus, it makes sense to replace one of the $I_N$ ($N=1-3$) with $n$ which now becomes one of 
the required invariants. Then we define $s^2$ to be a function of two of the other invariants. We 
can choose different combinations, but we must ensure that $s^2$ vanishes for the relaxed state. 
For example, Karlovini and Samuelsson \cite{Karlovini03:_elas_ns_1} work with
\begin{equation}
s^2 = {1\over 36} \left( I_1^3- I_3-24 \right) \ .
\label{Lars}
\end{equation}
In the limit $\eta_{ab} \to h_{ab}$ we have $I_1 , I_3 \to 3$ and therefore the combination for 
$s^2$ in Eq.~\eqref{Lars} vanishes.

Next, we assume that the Lagrangian of the system depends on $s^2$, rather than the tensor 
$k_{ab}$. In doing this, we need to keep in mind that Eqs.~\eqref{etadef} and \eqref{invs} show 
that the invariants $I_N$  depend on $n$ (and hence both $n^a$ and $g_{ab}$) as well as 
$k_{ab}$.

So far, the description is nonlinear, but in most situations of astrophysical interest it should be 
sufficient to consider a slightly deformed configuration\footnote{Note that this assumption is distinct from that of linear perturbations describing the dynamics.}. Then we may focus on a Hookean model, 
such that
\begin{equation}
\Lambda = - \check \varepsilon(n) - \check \mu(n) s^2 = - \varepsilon \ ,
\label{hooke}
\end{equation}
where $\check\mu$ is the shear modulus (not to be confused with the chemical potential). 
As mentioned earlier, the checks indicate that quantities are calculated for the 
unstrained state, with the specific understanding that 
$s^2=0$, and it should be apparent from \eqref{hooke} that we have an expansion in a 
supposedly small $s^2$. Since the strain scalar is given in terms of invariants, as in \eqref{Lars}, 
it might be tempting to suggest a change of variables such that $s^2=s^2(I_1,I_3)$. Our final 
equations of motion will, indeed, reflect this, but it would be premature to make the change at this 
point. 

Instead we note that we now have for the momentum
\begin{equation}
\mu_a = {\partial \Lambda \over \partial n^a} = {\partial n^2 \over \partial n^a} {\partial \Lambda\over  \partial n^2} = - {1 \over n} {\partial \Lambda\over  \partial n} g_{ab}n^b
=  {1\over n} \left( {d \check \varepsilon \over dn} + {d\check \mu\over dn} s^2 + \check \mu {\partial s^2 \over \partial n} \right) g_{ab}n^b \ ,
\end{equation}
while
\begin{equation}
{\partial \Lambda \over \partial g_{ab} }=-  \left( {d\check \varepsilon \over dn} + {d\check \mu\over dn} s^2 + \check \mu  {\partial s^2 \over \partial n} \right) {\partial n \over \partial g_{ab}} - \check \mu 
 {\partial s^2 \over \partial g_{ab}} \ .
 \label{Lambder}
\end{equation}
Here we need (note that $n^a$ is held fixed in the partial derivative)
\begin{equation}
 {\partial n \over \partial g_{ab}} = - {1\over 2n} n^a n^b  \ ,
\end{equation}
and it is useful to note that 
\begin{equation}
 {\partial s^2 \over \partial g_{ab}} = - g^{ad} g^{be} {\partial s^2 \over \partial g^{de}} \ .
\end{equation}
Also, when working out this derivative, we need to hold $n$ fixed [as is clear from \eqref{Lambder}].  
At the end of the day, we have for the stress-energy tensor
\begin{multline}
T^{ab} = \left[ \Lambda + n \left( {d\ec \over dn} + {d\check \mu\over dn} s^2 + \check \mu {\partial s^2 \over \partial n} \right)\right] g^{a b}   
    + {1\over n} \left( {d\ec \over dn} + {d\check \mu\over dn} s^2 + \check \mu  {\partial s^2 \over \partial n} \right) n^a n^b  +2  \check \mu g^{ad} g^{be} {\partial s^2 \over \partial g^{de}}
    \\
    =  \Lambda g^{ab} + n \left( {d\ec \over dn} + {d\check \mu\over dn} s^2 + \check \mu {\partial s^2 \over \partial n} \right) h^{ab}  +  2 \check \mu g^{ad} g^{be} {\partial s^2 \over \partial g^{de}} \ .
    \label{stress2}
\end{multline}

Let us now effect the change of variables we hinted at previously. To be specific, let us consider a 
situation where $s^2$ depends only on $I_1$. Then we need
\begin{equation} 
I_1 = \eta^a_{\ a} = n^{-2/3} g^{ab} k_{ab} \ ,
\end{equation}
\begin{equation}
\left({\partial s^2 \over \partial n}\right)_1= - {2I_1 \over 3n}  { \partial s^2 \over \partial I_1 } \ ,
\end{equation} 
\begin{equation}
\left({\partial \Lambda \over \partial k_{ab}}\right)_1 = - \check \mu 
{\partial s^2 \over \partial k_{ab} } = - \check \mu n^{-2/3} g^{ab} { \partial s^2 \over \partial I_1 } 
\  ,
\end{equation}
(recall comment on the partial derivative from before) and
\begin{equation}
\left({\partial s^2 \over \partial g^{de}}\right)_1 =  { \partial s^2 \over \partial I_1 }\eta_{de} \ .
\end{equation}
Making use of these results, we readily find
\begin{multline}
T^{ab} =  -\varepsilon g^{ab} + n \left( {d\ec \over dn} + {d\check \mu\over dn} s^2 \right) h^{ab} + 2  \check \mu { \partial s^2 \over \partial I_1 }  \left( \eta^{ab} - {1\over 3} I_1 h^{ab} \right)   \\
=  -\varepsilon g^{ab} + n \left( {d\ec \over dn} + {d\check \mu\over dn} s^2 \right) h^{ab} + 2  \check \mu { \partial s^2 \over \partial I_1 }  \eta^{\langle ab \rangle} \ , \label{stress3}
\end{multline}
where the $\langle \ldots \rangle$ brackets indicate the symmetric, trace-free part of a tensor with 
two free indices. In our case, we have
\beq
\eta_{\langle ab \rangle} = \eta_{(ab)} - { 1 \over 3} \eta^d_{\ d} h_{ab} \ . \label{angdef}
\eeq

Comparing this result to the standard decomposition of the stress-energy tensor, 
\beq
T^{ab} = \varepsilon u^a u^b + \bar p h^{ab} + \pi^{ab}\ , \qquad \mbox{where} \qquad \pi^a_{\ a} = 0 \ ,
\label{stress4}
\eeq
and $\bar p$ is the isotropic pressure (which differs from the fluid pressure, $p$, as it accounts 
for the elastic contribution, see below). We see that  elasticity introduces an anisotropic 
contribution
\begin{equation}
\pi^1_{ab} = 2 \check \mu {\partial s^2 \over \partial I_1}  \eta_{\langle ab \rangle}  \ .
\end{equation}

A similar analysis for the other two invariants, $I_2$ and $I_3$, leads to
\begin{equation}
I_2 = n^{-4/3} g^{ad} g^{be} k_{ea} k_{db} \ , 
\end{equation}
\begin{equation}
\left({\partial s^2 \over \partial n}\right)_2 = - {4 I_2 \over 3 n}  { \partial s^2 \over \partial I_2 } \ , 
\end{equation} 
\begin{equation}
\left({\partial s^2 \over \partial g^{de}}\right)_2 = 2 { \partial s^2 \over \partial I_2 } \eta^f{}_d 
\eta_{e f} 
\end{equation}
\begin{equation}
\left({\partial \Lambda \over \partial k_{ab}}\right)_2 = - 2 \check \mu n^{-4/3} k^{ab} { \partial s^2 \over \partial I_2 } \  , 
\end{equation}
and
\begin{equation}
I_3 = \eta^a_{\ a} = n^{-6/3} g^{ae} g^{bf} g^{dg} k_{ga} k_{eb} k_{fd} \ ,
\end{equation}
\begin{equation}
\left({\partial s^2 \over \partial n}\right)_3 = - {6 I_3 \over 3n}  { \partial s^2 \over \partial I_3 } \ ,
\end{equation} 
\begin{equation}
\left({\partial s^2 \over \partial g^{de}}\right)_3 =  3 { \partial s^2 \over \partial I_3 }  \eta^f{}_d 
\eta_{e g} \eta^g{}_f \ , 
\end{equation}
\begin{equation}
\left({\partial \Lambda \over \partial k_{ab}}\right)_3 = - 3 \check \mu n^{-6/3} k^{d a} k^b{}_d { \partial s^2 \over \partial I_3 } \  .
\end{equation}

Recalling the definition in Eq.~\eqref{angdef}, these lead to
\begin{equation}
\pi^2_{ab} = 4 \check \mu {\partial s^2 \over \partial I_2}   \eta_{d \langle a} \eta_{b \rangle}^{\ d} \ ,
\end{equation}
and
\begin{equation}
\pi^3_{ab}  = 6 \check \mu {\partial s^2 \over \partial I_3}  \eta^{d e} \eta_{d \langle a} \eta_{b \rangle e} \ , 
\end{equation}
respectively.
Combining these results with \eqref{Lars}, we have
\begin{equation}
\pi_{ab} = \sum_N \pi^N_{ab} =  {\check\mu\over 6}  \left[ \left(\eta^d_{\ d}\right)^2 \eta_{\langle ab\rangle}-  \eta^{d e} \eta_{d \langle a} \eta_{b\rangle e}\right] \ ,
\label{piab}
\end{equation}
which agrees with equation (128) from \cite{Karlovini03:_elas_ns_1}. 

Now consider the final stress-energy tensor. Note first of all that, if we consider $n$ and $s^2$ as 
the independent variables of the energy functional, then the isotropic pressure should follow from
\beq
\bar p = n \left( {\partial \varepsilon \over \partial n} \right)_{s^2} - \varepsilon = \check p + \left( \frac{n}{\check \mu} {d\check \mu\over dn} -1 \right) {\check \mu} s^2 \ ,
\eeq
where
\beq
\check p = n {d\ec \over dn}  - \ec  \ , 
\eeq
 is identical to the fluid pressure from before. However, we may also introduce a corresponding momentum, such that
\beq
\bar \mu_a = - \left( {\partial \Lambda \over \partial n^a} \right)_{s^2} = \left( {d\ec \over dn} + {d\check \mu\over dn} s^2 \right) n_a\ , \label{elmom}
\eeq
which  leads to 
\beq
\bar p = \Lambda - n^a \bar \mu_a =  \pc + \left( {n \over \muc} {d \muc \over dn} - 1 \right) \muc s^2 \ .
\eeq

Finally, in order to obtain the equations of motion for the system we can either take the 
divergence of \eqref{stress4} or return to \eqref{tforce} and make use of our various definitions. 
The results are the same (as they have to be). After a little bit of work we find that \eqref{tforce} 
leads to
\beq
2n^b\nabla_{[b}\bar \mu_{a]} + h_a^d \left( \nabla^b \pi_{b d} - \muc \nabla_d s^2\right) = 0 \ .
\label{finalmom}
\eeq
where it is worth noting that the combination in the parentheses is automatically flow line 
orthogonal. 

\section{Lagrangian perturbations of an unstrained medium}

The results in the previous section prepare the ground for a discussion of  Lagrangian 
perturbations of elastic matter.  In fact, we have already done most of the required work. In 
particular, we already know that 
\beq
  \Delta k_{ab} = 0 \ .
\label{delkab}\eeq
We now want to make maximal use of this fact. 

If we assume that the background configuration is relaxed, i.e. that $s^2=0$ vanishes for the 
configuration we are perturbing with respect to, then the fluid results from Section~\ref{langperb} 
together with \eqref{delkab} make the elastic perturbation problem straightforward (although it 
still involves a fair bit of algebra). 

Consider, first of all, the strain scalar. A few simple steps leads to
\beq
\Delta s^2 = 0 \ .
\label{dels2}\eeq
To see this, recall that $s^2$ is a function of the invariants, $I_N$. Express these in terms of the 
number density $n$, the spacetime metric and  $k_{ab}$. Once this is done, make use of  
\eqref{delkab} and the fact that the background is unstrained, i.e. $\eta_{ab} = h_{ab}$, to see 
that $\Delta I_N=0$. Intuitively, this result makes sense. Since the strain scalar is quadratic,  
linear perturbations away from a relaxed configuration should vanish. An important implication of 
this result is that the last term in \eqref{finalmom} does not contribute to the perturbed equations 
of motion.  

This strategy leads to
\beq
\Delta \eta_{ab} = {1\over 3} \eta_{ab} h^{de} \Delta g_{de}
\eeq
and 
\beq
\Delta \eta^{ab} = \left[ - 2 g^{a(e} \eta^{d)b} + {1\over 3} \eta^{ab} h^{de} \right] \Delta g_{de}
\eeq
It then follows from \eqref{sab} and \eqref{piab}, that
\beq
\Delta \pi_{ab} = - 2 \muc \Delta s_{ab} \ ,  
\eeq
where 
\beq
 2 \Delta s_{ab} = \left( h^e_{\ a} h^d_{\ b}  -  \frac13 h_{ab} h^{de}  \right) \Delta g_{de}  \ .
\eeq

It is worth noting that the final result for an isotropic material agrees with, for example, \cite{1983MNRAS.203..457S} where the relevant strain term is simply added to the stress-energy tensor (without particular justification). 

Finally, let us turn to the perturbed equations of motion. In the case of an unstrained background, 
it is easy to see that the argument that led to \eqref{euler1} still holds. This gives us the 
perturbation of the first term in \eqref{finalmom} (after replacing $\mu_a\to \bar \mu_a$). 
Similarly, since $\pi_{ab}$ vanishes in the background, the Lagrangian variation commutes with 
the covariant derivative in the second term. Thus, we end up with a perturbation equation of form
\beq
 2n^a\nabla_{[a} \Delta \bar \mu_{b]} + \nabla^a \Delta \pi_{ab} = 0  \ .
\eeq
This is the final result, but
in order to arrive at an explicit expression for the perturbed momentum, it is useful to note that
\beq
\Delta \mu_a = - {1 \over 2n} \betac u_a h^{{b} d} \Delta g_{{b} d} + \mu \left( \delta_a^{{b}} u^d
+ { 1 \over 2} u_a u^{{b}} u^d \right) \Delta g_{{b} d} \ , 
\eeq
where we have defined the bulk modulus $\betac$ as
\beq
\betac = n {d\pc \over dn} = (\pc + \check\varepsilon) {d\pc \over d \check\varepsilon} =  (\pc + \check \varepsilon) \csc \ ,
\eeq
$\csc$ is the sound speed in the elastic medium and we have also used the fundamental 
relation $\pc + \check \varepsilon = n \mu$.  It also follows that 
\beq
\Delta p = - {\check \beta \over 2} h^{ab} \Delta g_{ab} \ .
\eeq

When we consider perturbations of an elastic medium we need to pay careful attention to the 
magnitude of the deviation away from the relaxed state. If the perturbation is too large,  the 
material will yield \cite{2009PhRvL.102s1102H}.  It may fracture or behave in some other fashion 
that is not appropriately described by the equations of perfect elasticity. We need to quantify the 
associated breaking strain. In applications involving neutron stars, this is important if we want to 
consider star quakes in a spinning down pulsar, establish to what extent crust quakes in a 
magnetar lead to the observed flares \cite{Watts06:_xray_oscs} and whether the crust breaks 
due to the tidal interaction in an inspiralling binary 
\cite{2012ApJ...749L..36P,2012PhRvL.108a1102T}.

A  commonly used criterion to discuss elastic yield strains in engineering involves the von Mises 
stress, defined as
\beq\label{vMdef}
  \Theta_{\mathrm{vM}} = \sqrt{\frac32 s_{ab}s^{ab}}
\eeq
When this scalar exceeds some critical value 
$\Theta_{\mathrm{vM}} > \Theta^{\mathrm{crit}}_{\mathrm{vM}}$, say, the material no longer 
behaves elastically and the framework we have developed needs to be amended. In order to 
work out the dominant contribution to the von Mises stress in general we need to (at least 
formally) consider second order perturbation theory. This is due to the positive definite nature of 
\eqref{vMdef} which implies that the first order perturbation is zero for unstrained backgrounds. 
We could perturb \eqref{vMdef} directly to second order, but it turns out to be simpler (and more 
elegant) to expand the trace of the squared strain tensor separately and then calculate the von 
Mises stress. This works because the von Mises stress is not a primary variable needed to solve 
the perturbation equations, but rather a quantity that can be estimated in post-processing.

Hgher-order perturbations, when the strain  $s_{a b}$ is considered ``small'' in the sense that $\Theta_{\mathrm{vM}} << \Theta^{\mathrm{crit}}_{\mathrm{vM}}$, then formally  involve the substitution 
\begin{align}
   s_{ab} &\rightarrow s_{ab} + \Delta s_{ab} + \Delta^{(2)} s_{ab} + \ldots \\
   s^{ab} &\rightarrow s^{ab} + \Delta s^{ab} + \Delta^{(2)} s^{ab} + \ldots \ .
\end{align}
This leads to the trace of the squared strain tensor 
\beq
   s_{ab}s^{ab} \rightarrow s_{ab}s^{ab} + (s_{ab}\Delta s^{ab} + s^{ab}\Delta s_{ab}) + (\Delta s_{ab}\Delta 
   s^{ab} + s_{ab}\Delta^{(2)} s^{ab} + s^{ab}\Delta^{(2)} s_{ab}) + \ldots
\eeq
where we have grouped the terms according to the perturbative order. Although we cannot say 
anything about the relative size of $s_{ab}$ and $\Delta s_{ab}$ (this involves choice), we do 
know that $\Delta^{(2)} s_{ab} \ll \Delta s_{ab}$ (and similar for the contravariant stress tensor) so  
the last two terms can be neglected compared to the linear order terms. Making use of this, we 
have
\begin{multline}
  \Theta_{\mathrm{vM}} \approx  \sqrt{\frac32 s_{ab}s^{ab} + \frac32(s^{\langle ab \rangle} - 2 s^a{}_c s^{cb})\Delta g_{ab} + \frac38 h^{a\langle c}h^{d\rangle b}\Delta g_{ab}\Delta g_{cd}} \\
         \approx  \sqrt{\frac32 s_{ab}s^{ab} + \frac32 s^{\langle ab \rangle}\Delta g_{ab} + \frac38 h^{a\langle c}h^{d\rangle b}\Delta g_{ab}\Delta g_{cd}}
         \label{vM3}
\end{multline}
where the smallness of the strain tensor was used in the last step.

The relevant comparison is between the size of the background strain tensor and the spatially 
projected trace-free part of the perturbed metric. If either is much larger than the other, then 
\eqref{vM3} simplifies. For instance, if the background is unstrained (or weakly strained) we have  
\beq\label{vM1}
  \Theta_{\mathrm{vM}} = \sqrt{\frac32 \Delta s_{ab} \Delta s^{ab}} 
     = \sqrt{\frac38 h^{a\langle c}h^{d\rangle b}\Delta g_{ab}\Delta g_{cd}}
\eeq
A very neat result, indeed. If, on the other hand, we consider perturbations of a strained background 
the relevant expression is
\begin{align}\label{vM2}
  \Theta_{\mathrm{vM}} &\approx \sqrt{\frac32\left(s^{ab}s_{ab} + s^{\langle ab\rangle }\Delta g_{ab}\right)}
\end{align}
As an aside, it is worth noting that this expression neatly demonstrates the interpretation of 
perturbed spacetime as a strain.

\section{Adding an entrained superfluid component}

In order to develop a model for the coupled superfluid neutrons-elastic nucleon crust
we need to account for the presence of two (coupled) fluxes\footnote{We are assuming that the 
second fluid contribution represents superfluid neutrons, but from a formal point of view it could 
equally well correspond to a dynamical thermal component \cite{andersson07:_livrev}.}. We take 
these to be $n_\I^a$ and $n_\f^a$, where the constituent indices $\x=\I$ and $\f$ distinguish the 
``confined'' baryons in the lattice from the ``free'' (superfluid) neutrons\footnote{The careful 
reader will  note that, in order to avoid confusion  we have left out the letters $c$ and $f$ as 
spacetime indices from this point.}. Now we have two distinct four velocities, such that 
$n_\I^a=n_\I u_\I^a$ and $n_\f^a=n_\f u_\f^a$. The simplest relevant model for the matter 
Lagrangian of this system assumes that the elastic contribution is unaffected by the presence of 
the interpenetrating fluid component. Assuming a Hookean model, we then have [c.f. 
\eqref{hooke}]
\beq
\Lambda = \Lambda_\mathrm{liq} (n_\I^a, n_\f^a , g_{a b})   + \Lambda_\mathrm{sol}  
\left(n_\I , s^2 \right)
\eeq
where we have made a ``minimal coupling'' assumption for the elastic contribution, 
$\Lambda_\mathrm{sol}$. That is,  the corresponding matter component is associated with the 
constituent index $\I$, such that $\muc=\muc(n_\I)$. The liquid contribution is, of course, different 
from before. In general, $\Lambda_\mathrm{liq}$ is  a function of three scalar densities 
\cite{andersson07:_livrev}:
\beq
n_\f^2 = - n^\f_a n_\f^a \ ,
\quad
n_\I^2 = - n^\I_a n_\I^a \ ,
\quad
n_{\f\I}^2 = - n^\f_a n_\I^a \ .
\eeq
The last of these represents effects due to the relative flow between the two components. While 
this flow is generally expected to be small in magnitude, its contribution is nevertheless 
significant since it encodes the entrainment effect 
\cite{andreev75:_three_velocity_hydro,borumand96:_superfl_neutr_star_matter,comer03:_rel_ent,carter06:_ent_nscrust}.

At this point it makes sense to point out that, in the neutron star crust, the distinction 
between the two dynamical components is somewhat ambiguous. Throughout most of the crust 
(beyond neutron drip),  we have a fraction of neutrons bound in nuclei but there is also a ``gas'' 
of free neutrons. In static situations, neutrons can be assigned to either component depending on 
the nature of the ions in the lattice \cite{Chamel08:_LRR}. However, when we turn to dynamical 
settings,  it is no longer clear to what extent the ``confined'' neutrons are able to move 
\cite{carter06:_crust,Chamel12:_neut_con,Chamel13:_free_neu}. The answer depends on the 
extent to which they can tunnel through the relevant interaction potentials, an effect that can be 
expressed in terms of the entrainment. As a result, while it is clear that we must deal with a 
two-component model, it is conceptually less clear how one  determines the parameters of the 
system.  Formally, one may consider the problem in terms of different chemical ``gauges'' 
\cite{carter06:_crust}. This is tricky.  Fortunately, while the choice of chemical gauge affects the 
interpretation of the involved quantities (number densities, etc),  the two-fluid model remains 
conceptually unaffected \cite{2011MNRAS.416..118A}. The upshot is that one has to exercise a 
level of care in practical applications where the model is combined with a detailed microphysical 
equation of state.

\subsection{The background dynamics}

Given the form of $\Lambda$, the variational procedure determines the momenta that are 
canonically conjugate to the two fluxes. We have
\beq
  \mu_a^\f = \frac{\partial \Lambda}{\partial n^a_\f}
           = \frac{\partial \Lambda_{\mathrm{liq}}}{\partial n^a_\f} \ , \qquad
  \mu_a^\I = \frac{\partial \Lambda}{\partial n^a_\I} \ .
\eeq
and the stress-energy tensor takes the form
\beq
  T^a{}_{b} = (\Lambda - n_\f^d\mu_d^\f -  n_\I^d\mu_d^\I)\delta^a{}_{b}
     +  n_\f^a\mu_b^\f + n_\I^a\mu_b^\I + \pi^a{}_b \ ,
\eeq
where the anisotropic pressure, $\pi_{ab}$, is still given by \eqref{piab}. We also need the 
generalized pressure
\beq
  \Psi = \Lambda - n_\I^a\mu_a^\I - n_\f^a\mu_a^\f \ .
\eeq

The stress-energy tensor serves as source for Einstein's equations. Moreover, in the case where 
the two fluxes are individually conserved, i.e., when we are not accounting for reactions (for 
example, when the dynamical timescale is much faster than that associated with reactions), we 
have
\beq
\nabla_a n_\I^a =
\nabla_a n_\f^a = 0 \ .
\eeq
In this case, we obtain two
equations of motion
\begin{align}
  2n^a_\f\nabla_{[a}\mu^\f_{b]} &= 0 \ ,\label{eulerf}\\
  2n^a_\I\nabla_{[a}\mu^\I_{b]} + \nabla^a\pi_{ab} &= 0  \ .\label{eulerc}
\end{align}
Given the previous discussion, the form of these equations should come as no surprise.

It should be noted that Equations \eqref{eulerf} 
and \eqref{eulerc}, which represent the Euler equations, combine to ensure the conservation of 
energy momentum, $\nabla^aT_a^b = 0$. This information is, of course, also encoded in the 
Einstein equations. Thus, it is sufficient to consider a combination of the Einstein equations and 
one of the Euler equations. An often used strategy, especially in work on neutron star oscillations 
is to focus on the Einstein equations which, for a single component fluid, contain all required 
information. In the two-fluid case this strategy will not completely specify the problem 
\cite{comer99:_quasimodes_sf,andersson02:_oscil_GR_superfl_NS}. We also 
need information from \eqref{eulerf} and/or \eqref{eulerc}. From the formal point of view, the 
tidiest approach may be to use both Euler equations and a smaller subset of the Einstein field 
equations. A key reason for this is that one can then develop the model in such a way that many 
of the equations are ``symmetric'' in the constituent indices. This makes the description  
economical, and  has the advantage that the inclusion of additional fluid components is 
straightforward. Of course, it comes at the price of having to work with the constituent indices at 
a more abstract level.

To complete the model, and obtain explicit equations, we need to determine the fluid momenta. 
Adapting the notation from \cite{andersson07:_livrev} we have
\begin{align}
  \mu_a^\f &= \Bcal^\f n_a^\f + \Acal^{\I\f} n_a^\I \label{muf}\\
  \mu_a^\I &= \Bcal^\I n_a^\I + \Acal^{\I\f} n_a^\f \label{muc}
\end{align}
where
\beq
  \Bcal^\f     = -2\frac{\partial \Lambda}{\partial n_\f^2} \ , \quad
  \Bcal^\I     = -2\frac{\partial \Lambda}{\partial n_\I^2} \ , \quad
  \Acal^{\I\f} = -\frac{\partial \Lambda}{\partial n_{\I\f}^2} \ .
\eeq

\subsection{The perturbation equations}

Let us now turn to the Lagrangian perturbations of this system. In principle, this problem is 
straightforward given the previous developments. It is natural to work with two matter 
spaces \cite{andersson07:_livrev} and, hence, two distinct displacements $\xi_\f^a$ and 
$\xi_\I^a$  (see \cite{andersson04:_canon_energy} for a discussion of the corresponding 
Newtonian problem). 

From the single-fluid results in Section~\ref{langperb}, it is easy to see that we will have 
\beq
\Delta_\x n_\x^a = - n_\x^a \left( \nabla_b \xi_\x^b + { 1 \over 2} g^{bd} \delta g_{bd} \right) = - {1 \over 2} n_\x^a \left( g^{bd} \Delta_\x g_{bd}\right)\ ,
\label{dnax}\eeq
where  the constituent index $\x$ represents either $\f$ or $\I$, and
\beq
\Delta_\x g_{ab} = \delta g_{ab} + 2\nabla_{(a}\xi^\x_{b)} \ .
\eeq
This leads to
\beq
\Delta_\x n_\x = -{ n_\x \over 2} h_\x^{ab} \Delta_\x g_{ab} \ ,
\eeq
where
\beq
h_\x^{ab} = g^{ab} + u_\x^a u_\x^b \ ,
\eeq
is the projection orthogonal to $u_\x^a$. We also get
\beq
\Delta_\x u_\x^a = { 1 \over 2} u_\x^a u_\x^b u_\x^d \Delta_\x g_{bd} \ .
\eeq
Moreover, the argument that led to the perturbed equations of motion remains valid (as long as 
we assume that the elastic background is isotropic and unstrained) and we have
\begin{align}
  2n^a_\f \nabla_{[a} \Delta_\f \mu^\f_{b]} &= 0 \ ,\label{peulerf}\\
  2n^a_\I \nabla_{[a} \Delta_\I \mu^\I_{b]} + \nabla^a \Delta_\I \pi_{ab} &= 0  \ .\label{peulerc}
\end{align}
These are the main results. Of course, given the two-fluid context, the perturbed momenta are 
more complicated than before.

Starting from \eqref{muf}--\eqref{muc}, one can show that \cite{andersson07:_livrev}
\begin{multline}
\Delta_\x \mu_a^\x = \left( \Bcal^\x_{ab} + \Acal^{\x\x}_{ab} \right) \Delta_\x n_\x^b + 
\left( \chi^{\x\y}_{ab} + \Acal^{\x\y}_{ab} \right) \Delta_\x n_\y^b \\
+ {1 \over 2} g^{db} \left[ \delta^e_a \mu_b^\x + \left( \Bcal^\x_{ab} + \Acal^{\x\x}_{ab} \right) n_\x^e + \left( \chi_{ab}^{\x\y}+\Acal^{\x\y}_{ab}  \right)n_\y^e \right] \Delta_\x g_{ed}  \ ,
\label{dmux}\end{multline}
where we have introduced another constituent index, $\y \neq \x$.
We have also defined
\beq
\Bcal_{ab}^\x = \Bcal^\x g_{ab} - 2 {\partial \Bcal^\x \over \partial n_\x^2} n^\x_a n^\x_b \ , 
\eeq
in obvious analogy with \eqref{Bab}, 
\beq
\chi_{ab}^{\x\y} = - 2  {\partial \Bcal^\x \over \partial n_\y^2} n^\x_a n^\y_b \ , 
\eeq
\beq
\Acal_{ab}^{\x\x} = -  {\partial \Bcal^\x \over \partial n_{\x\y}^2} \left( n^\x_a n^\y_b + n^\x_b n^\y_a \right) - {\partial \Acal^{\x\y} \over \partial  n_{\x\y}^2} n^\y_a n^\y_b \ ,
\eeq
and
\beq
\Acal^{\x\y}_{ab} = \Acal^{\x\y} g_{ab} - {\partial \Bcal^\x \over \partial n_{\x\y}^2} n^\x_a n^\x_b - {\partial \Bcal^\y \over \partial n_{\x\y}^2} n^\y_a n^\y_b 
- {\partial \Acal^{\x\y} \over \partial  n_{\x\y}^2} n^\y_a n^\x_b \ .
\eeq
From these expressions, it is apparent that we also need 
\beq
\Delta_\x n_\y^a = \Delta_\y n_\y^a +  (\xi_\x^b - \xi_\y^b) \nabla_b n_\y^a - n_\y^b \nabla_b (\xi_\x^a- \xi_\y^a)   \ , 
\label{single}\eeq
and it is useful to note that 
\beq
\Delta_\y g_{ab} = \Delta_\x g_{ab} - 2 \left[ \nabla_{(a}\xi^{\x}_{b)}- \nabla_{(a}\xi^{\y}_{b)}\right]\ .
\eeq

These relations provide all the information we need in order to express the perturbations in terms 
of the two displacement vectors $\xi_\x^a$ and the perturbed metric $\delta g_{ab}$. Once an 
equation of state is provided (so that we can work out the action), all required coefficients 
can be calculated and we have a description of a generic situation.  

\subsection{A couple of steps towards applications}

Even though our main aim is to establish formal aspects of the Lagrangian perturbation problem, 
it makes sense to make a few comments on applications.  In particular, it is worth noting that the 
magnitude of the relative velocity between the two components is likely to be small in most 
situations of practical relevance.  

In order to quantify the relative flow, let us focus on the frame associated with one of the fluids. 
Taking the four velocity $u^a=u^a_\I$ as our reference, the relative velocity $v^a$ follows from
\beq\label{nfproj}
  n_\f^a = \gamma n_\f(u^a + v^a) \ ,
  \qquad u^av_a = 0 \ ,
  \qquad \gamma = (1-v^2)^{-1/2} \ ,
  \qquad v^2 = v^av_a \ .
\eeq
Assuming that $v^2$ is small, it makes sense to work with an expansion using this as a small 
parameter.

With this in mind, and considering the variables that we used in the derivation of the equations of 
motion, we may follow \cite{comer04:_rel_ent_slowrot,2009CQGra..26o5016S} and expand the Lagrangian as 
\beq
\Lambda(n_\f^2, n_\I^2, n_{\f\I}^2) \approx \sum_{i=0}^N \lambda_i(n_\f^2,n_\I^2) \left[ n_{\f\I}^2 - n_\f n_\I \right]^i \ .
\label{lambexp}
\eeq
Since
\beq
n_{\f\I}^2 - n_\f n_\I \approx { 1 \over 2} n_\f n_\I v^2 \ ,
\eeq
it should be sufficient to retain the first couple of terms in the expansion. For example, if we 
accept errors of order $v^2$ in the equations of motion then we need to keep the first three 
terms, up to $N=2$. At this level of precision, we get
\beq
\Bcal^\f \approx - { 1 \over n_\f} {\partial \lambda_0 \over \partial n_\f} + { n_\I \over n_\f } \lambda_1 \ ,
\eeq
\beq 
\Bcal^\I\approx - { 1 \over n_\I} {\partial \lambda_0 \over \partial n_\I} + { n_\f \over n_\I } \lambda_1 \ ,
\eeq
and
\beq
\Acal^{\f\I} \approx - \lambda_1 \ .
\eeq

To completely specify the model, we  need to provide the $\lambda_i$ coefficients. In 
order to illustrate how different features enter at different levels of complexity, we can start by considering models where the two fluids co-move in the background. This will be the case if one insists that  the background configuration is in both dynamical and chemical equilibrium.
 
 If we take the two fluids to move together in the background we have 
$u^a_\f = u^a_\I = u^a$ and the problem simplifies. The Lagrangian is simply 
given by $\Lambda = \lambda_0$ and it makes sense (as in the single-fluid problem) to work with 
the energy density $\rhoc=-\lambda_0$. We also find that the pressure is given by 
\beq
\pc = - \rhoc + n_\N \mu_\N + n_\n \mu_\n = - \rhoc + n_\N { \partial \rhoc \over \partial n_\N} + n_\n { \partial \rhoc \over \partial n_\n} \ .
\label{fundam}\eeq

Combining this with \eqref{dmux}, we find that the perturbed momenta can be written
\begin{multline}
\Delta_\x \mu^\x_a = -  { 1 \over 2} \left( n_\x {\partial^2 \rhoc \over \partial n_\x^2} + n_\y  {\partial^2 \rhoc \over \partial n_\x \partial n_\y} \right) u_a \perp^{de} \Delta_\x g_{de}
+ {\partial \rhoc \over \partial n_\x} \left( \delta^e_a u^d + {1 \over 2} u_a u^e u^d \right) \Delta_\x g_{ed} 
\\
- {\partial^2 \rhoc \over \partial n_\x \partial n_\y} \left[ \psi_{\x\y}^d \nabla_d n_\y - n_\y \perp^{cd} \nabla_c \psi_d^{\x\y}  \right]  u_a  \ , \qquad \y \neq \x \ .
\end{multline}
Alternatively, introducing 
\beq
\betac_\x = n_\x \mu_\x = n_\x { \partial \pc \over \partial n_\x} \ , 
\eeq
we have 
\begin{multline}
\Delta_\x \mu^\x_a = -  { \betac_\x \over 2 n_\x} u_a \perp^{de} \Delta_\x g_{de}
+ \mu_\x \left( \delta^e_a u^d + {1 \over 2} u_a u^e u^d \right) \Delta_\x g_{ed}
\\
- {\partial \mu_\x \over \partial n_\y} \left( \psi_{\x\y}^d \nabla_d n_\y - n_\y \perp^{cd} \nabla_c \psi_d^{\x\y}  \right)  u_a  \ , \qquad \y \neq \x \ .
\label{dmu1}\end{multline}
The single-fluid result \eqref{single} provides a useful sanity check on this result. Comparing, we 
see that when the equation of state depends on the composition (and nuclear physics parameters like the symmetry energy), i.e. when 
\beq
{\partial \mu_\x \over \partial n_\y} \neq 0 \ , 
\eeq
there are two key differences. First of all, variations in the composition affect $\betac_\x$. In the 
case when the two components are coupled also at the perturbative level, this change leads to 
the presence of g-modes etcetera \cite{reisenegger92}. Secondly, when the two fluids are free to move relative 
one another, the associated displacements are ``chemically'' coupled through the last two terms 
in \eqref{dmu1}.  

The problem gets significantly more involved if the two fluids are not flowing together in the 
unperturbed configuration. The general expression for the perturbed momenta, \eqref{dmux}, 
remains valid but the involved terms are more complex. Having said that, one should be 
able to neglect all quadratic terms in the relative velocity in most situations of practical interest.
To make the dependence on the relative velocity, $v^a$, explicit it may be useful express the 
perturbation equations in a prefered frame. However, this strategy breaks the ``symmetry'' with 
respect to the constituent indices that we have relied upon so far. 

Let us opt to work in the frame associated with the crust component, taking $u^a= u_\N^a$. To 
linear order in the relative velocity, we then have
$u_\f^a \approx u^a+v^a$. It  follows that the two unperturbed momenta are given by
\beq
\mu_a^\N \approx - { \partial \lambda_0 \over \partial n_\N} u_a - n_\n\lambda_1 v_a \ , 
\eeq
and
\beq
\mu_a^\n \approx - {\partial \lambda_0 \over \partial n_\n} (u_a + v_a) - n_\N\lambda_1 v_a \ , 
\eeq
Using these results, it is straightforward to show that (up to terms of order $v^2$) the energy 
density $\rhoc$ and the pressure $\pc$ remain as in the previous model problem. This tells us that 
the problem does not deviate too far from the co-moving situation as long as we neglect higher 
order terms in the relative velocity. 

In this model, the entrainment enters through the $\lambda_1$ coefficient. However, it is often useful to represent the effect in terms of an effective nucleon 
mass. This makes sense intuitively, and it also relates to a quantity that can be determined from 
detailed microphysics \cite{prix02:_slow_rot_ns_entrain}. In recent years, there has been an effort to determine the effective 
neutron mass for the crust superfluid. Perhaps surprisingly, this work \cite{carter06:_ent_nscrust,Chamel12:_neut_con} suggests that the 
effective mass may be very different from the bare nucleon mass. This would mean that the 
inclusion of entrainment in the treatment of the crust superfluid is essential \cite{2012PhRvL.109x1103A,Chamel13:_free_neu}.

Let us see how the effective mass arises in a relativistic model \cite{2009CQGra..26o5016S}. We can do this by 
considering the momentum in a local inertial frame associated with one of the fluids, say the 
crust component. Then we have
\beq
u_\N^a = [ 1, 0,0,0] \ , \quad u_\n^a = [ \gamma, \gamma v^a] \ ,
\eeq
with $\gamma = (1 - v^2)^{-1/2}$ as before.
This leads to
\beq
\mu_\n^0 = \B^\n n_\n \gamma + \A^{\n\N}n_\N = \gamma m_0 \ ,
\eeq
where $m_0$ is the baryon rest mass (we assume that $m_\n=m_\N=m_0$ here)
and
\beq
\mu_\n^i = \B^\n n_\n \gamma v^i \equiv m_\n^* \gamma v^i \ , \quad  i=1-3\ ,
\eeq
where $m_\n^*$ is the effective neutron mass. It follows that
\beq
\A^{\n\N} = { \gamma \over n_\N}  (m_0-m_\n^*) \ ,
\label{afc}
\eeq
 which reduces to the usual Newtonian result  \cite{prix02:_slow_rot_ns_entrain}
\beq
\A^{\n\N} = { 1 \over n_\N}  (m_0-m_\n^*) \ , 
\eeq 
in  the limit $v^2 \ll c^2$.

In order to make contact with the low-velocity expansion of the Lagrangian \eqref{lambexp}, we note that
\beq
\A^{\n\N} = - { \partial \Lambda \over \partial n_{\n\N}^2}
= - { 1 \over n_\n n_\N} {\partial \Lambda \over \partial \gamma} \ ,
\eeq
where 
\beq
\left. {\partial \Lambda \over \partial \gamma}\right|_{n_\n, n_\N} =  n_\n \gamma  (m_\n^*-m_0) \ .
\eeq
In general, one would expect the effective mass to depend on  $\gamma$, preventing us from 
integrating to get an expression for $\Lambda$.  Assuming that this dependence is weak 
enough that it can be ignored, we have
\beq
\Lambda = \Lambda_0(n_\n,n_\N) +  { 1 \over 2} n_\n (m_\n^*-m_0) \gamma^2 \ ,
\eeq
which leads to 
\beq
\lambda_1 = { n_\n ( m_\n^* - m_0) \over 2} \ .
\eeq

Finally, it is worth noting that we could (obviously) have introduced an analogous effective mass, 
$m_\N^*$, for the crust nucleons. Because of the symmetry of the entrainment terms, the two 
effective quantities must be related by
\beq
\A^{\n\N} = { 1 \over n_\N}  (m_0-m_\n^*) = { 1 \over n_\n}  (m_0-m_\N^*) \quad \longrightarrow m_\N^* = m_0 - {n_\n \over n_\N } \left( m_0 - m_\n^* \right) \ . 
\eeq

\section{Summary}

We have developed a Lagrangian perturbation framework for the dynamics of mature neutron 
stars with an  inner crust combining  an elastic lattice of neutron-rich nuclei and a free neutron 
component. We paid particular attention to geometric aspects of the problem and the close 
connection to the convective variational approach often used to derive the equations for 
multi-fluid systems in general relativity \cite{andersson07:_livrev}. As the final perturbation 
equations may be somewhat intimidating, we outlined simplifying assumptions that may apply to 
problems of astrophysical relevance. This discussion should also help build intuition. 

The natural next step will be to apply the result to a specific problem of interest. This may take us 
in different directions, as the formalism lends itself to a range of settings, from crust quakes \st{lead} 
to pulsar glitches to magnetar seismology and continuous gravitational-wave emission from 
rotating deformed neutron stars. The last problem is (perhaps) particularly timely given the 
excitement associated with the breakthrough detection of gravitational waves from neutron star 
mergers \cite{2017PhRvL.119p1101A} and the ongoing effort to detect gravitational waves from 
spinning neutron stars \cite{2017ApJ...839...12A}. We hope to make progress on this problem in 
the near future. 

\section*{Appendix: The properties of $k_{a b}$ and $\eta_{a b}$}

In this Appendix we add some relevant context relating to the description of elastic matter. In 
particular, we make use of a positive definite, matter-space metric tensor field 
$k_{A B} = k_{B A}$ to establish the major features of $k_{a b}$ and $\eta_{a b}$ required for the 
calculations presented in Sec.~\ref{relelas}. The tensor $k_{A B}$ is ``fixed'' on matter space, in 
the same sense as $n_{A B C}$, because it is only a function of its own matter space 
coordinates $X^A$. The associated volume form is $n_{A B C}$ in that 
\beq
      n_{A B C} = \sqrt{\det{\left(k_{A B}\right)}} \left[A \ B \ C\right]_{\cal D} \ , \label{kn3form}
\eeq
where
\beq
      \left[A \ B \ C\right]_{\cal D} = 3! \delta_{[A}^1 \delta_B^2 \delta_{C]}^3 = \{\pm 1, 0\} \ ,
\eeq
and 
\beq
      \det{\left(k_{A B}\right)} = \frac{1}{3!} \left[A \ B \ C\right]^{\cal U} \left[D \ E \ F\right]^{\cal U} 
      k_{A D} k_{B E} k_{C F} \ ,
\eeq
where
\beq
      \left[A \ B \ C\right]^{\cal U} = 3! \delta^{[A}_1 \delta^B_2 \delta^{C]}_3 = \{\pm 1, 0\} \ .
\eeq
Later it will be useful to know
\beq
      \left[A \ B \ C\right]^{\cal U} \left[D \ E \ F\right]_{\cal D} = 3! \delta^{[A}_D \delta^B_E 
                      \delta^{C]}_F \ . \label{altsymproduct}
\eeq
If we let
\beq
      g^{A B} = \psi^A_a \psi^B_b g^{a b} = \psi^A_a \psi^B_b h^{a b} \ ,
\eeq
and use Eqs.~\eqref{n3form} and \eqref{pb3form}, then we can show
\beq
     n^2 = - g_{a b} n^a n^b = \frac{1}{3!} \det{\left(k_{A B}\right)} \det{\left(g^{A B}\right)} \ , 
     \label{n2eqkg}
\eeq
where
\beq
       \det{\left(g^{A B}\right)} = \frac{1}{3!} \left[A \ B \ C\right]_{\cal D} \left[D \ E \ F\right]_{\cal D} 
       g^{A D} g^{B E} g^{C F} \ .
\eeq

Using Eqs.~\eqref{DelX} and \eqref{kAB}, we can easily establish that the Lagrangian variation of 
$k_{a b}$ vanishes; namely,
\beq
      \delta k_{a b} = - {\mathcal L}_\xi k_{a b} \quad \Longrightarrow \quad \Delta k_{a b} = 0 \ .
      \label{Delkab}
\eeq
Finally, since $u^a \psi^A_a = 0$, and $k_{A B}$ is a function of $X^A$, we have
\beq
      {\mathcal L}_u k_{A B} = u^a \psi^C_a \frac{\partial k_{A B}}{\partial X^C} = 0 \ ,
\eeq
and therefore
\begin{multline}
       {\mathcal L}_u k_{a b} = k_{A B} {\mathcal L}_u \psi^A_a \psi^B_b  \\
      = k_{A B} \left[u^c \frac{\partial}{\partial x^c} \left(\psi^A_a \psi^B_b\right) + \psi^A_c \psi^B_b \frac{\partial u^c}{\partial x^a} + \psi^A_a \psi^B_c \frac{\partial u^c}{\partial x^b}\right] \\
       = k_{A B} u^c \left[\frac{\partial^2 X^A}{\partial x^c \partial x^a} \psi^B_b + \psi^A_a \frac{\partial^2 X^B}{\partial x^c \partial x^b} - \frac{\partial^2 X^A}{\partial x^a \partial x^c} \psi^B_b - \psi^A_a \frac{\partial^2 X^B}{\partial x^b \partial x^c}\right] = 0 \ .
\end{multline}
In fact, one can prove that the Lagrangian variation of \underline{all} such tensors vanishes (see Carter and Quintana \cite{Carter72:_gr_elas}). 

Because $k_{a b}$ is flowline orthogonal, computing its determinant requires some work.  The 
key problem is that even though $k_{a b}$ carries full spacetime indices, it is degenerate, 
effectively meaning its full spacetime determinant is zero. So, let us consider the problem from 
the perspective of a local frame which follows the $u^a$ worldline. Clearly, $u^i = 0$ in this frame 
and we can choose the local time to match the proper time so that $u^0 = 1$. Finally, if we now 
consider a point on the worldline, we can arrange for the metric to be the flat-space metric in 
Minkowski coordinates. This means several things:
\begin{eqnarray}
       u^a \psi^A_a &=& \psi^A_0 = 0 \ , \\
       u^a k_{a b} &=& u^0 k_{0 b} + u^i k_{i b} = k_{0 b} = 0 \ , \\
       g^{A B} &=& \psi^A_1 \psi^B_1 + \psi^A_2 \psi^B_2 + \psi^A_2 \psi^B_2 \ ,
\end{eqnarray}
since $g^{1 1} = g^{2 2} = g^{3 3} =1$ and the off-diagonal components are zero.  Therefore, the 
only non-zero components of $k_{a b}$ are the $k_{i j}$ and the determinant of $k_{a b}$, to be 
denoted $k$, is obtained from
\begin{multline}
      k = \frac{1}{3!} \left(3! \delta^{[i}_1 \delta^j_2 \delta^{k]}_3\right) \left(3! \delta^{[l}_1 \delta^m_2 \delta^{n]}_3\right) k_{i l} k_{j m} k_{k n} \cr
        = \frac{1}{3!} \left(3! \delta^{[i}_1 \delta^j_2 \delta^{k]}_3\right) \left(3! \delta^{[l}_1 \delta^m_2 \delta^{n]}_3\right) \psi^{[A}_i \psi^B_j \psi^{C]}_k \psi^{[D}_l \psi^E_m \psi^{F]}_n k_{A D} k_{B E} k_{C F} \cr
         = \frac{1}{3!} \left(\psi^{[G}_1 \psi^H_2 \psi^{I]}_3 \delta^{[A}_G \delta^B_H \delta^{C]}_I\right) \left(\psi^{[J}_1 \psi^K_2 \psi^{L]}_3 \delta^{[D}_J \delta^E_K \delta^{F]}_L\right) k_{A D} k_{B E} k_{C F} \cr
         = \frac{1}{3!} \left(\frac{1}{3!} \left[G \ H \ I\right]_{\cal D} \left[J \ K \ L\right]_{\cal D} g^{G J} g^{H K} g^{I L}\right) \left(\frac{1}{3!} \left[A \ B \ C\right]^{\cal U} \left[D \ E \ F\right]^{\cal U} k_{A D} k_{B E} k_{C F}\right) \cr
         = \frac{1}{3!}  \det{\left(k_{A B}\right)} \det{\left(g^{A B}\right)} \ ,  
\end{multline}
where we have used Eq.~\eqref{altsymproduct}.  Upon comparing with Eq.~\eqref{n2eqkg} we see $k = n^2$.

Karlovini and Samuelsson \cite{Karlovini03:_elas_ns_1} introduce the matter space tensor 
$\eta_{A B}$ to quantify the so-called 
{\em unsheared} state.  Its defining characteristic is that it is the inverse to $g^{A B}$ but only for 
the unsheared state (when the energy density $\epsilon = \check{\epsilon}$):
\beq
       g^{A C} \eta_{C B} = \delta^A_B \quad , \quad \epsilon = \check{\epsilon} \ . \label{etainv}
\eeq 
If we introduce
\beq
      \epsilon^{A B C} = \psi^A_a \psi^B_b \psi^C_c u_d \epsilon^{d a b c} \ ,
\eeq
then from Eq.~\eqref{pb3form} we can infer
\beq
      n_{A B C} = n \epsilon_{A B C} \ ,
\eeq
where
\beq
      \epsilon^{A B C} \epsilon_{D E G} = 3! \delta^{[A}_D \delta^B_E \delta^{C]}_G \ ,
\eeq 
and
\beq
      \epsilon^{A B C} \epsilon^{D E F} \eta_{A D} \eta_{B E} \eta_{C F} = 3! \ ;
\eeq
in other words, 
\beq
      \epsilon_{A B C} = \sqrt{\det{\left(\eta_{A B}\right)}} \left[A \ B \ C\right]_{\cal D} \ .
\eeq
This tensor is useful because it allows a straightforward way to model conformal elastic 
deformations; namely, if $f$ is the conformal factor, we let
\beq
       k_{A B} = f \eta_{A B} \quad \Longrightarrow \quad  \det{\left(k_{A B}\right)} = f^3  \det{\left(\eta_{A B}\right)} \ .
\eeq
But,
\beq
      n_{A B C} = \sqrt{ \det{\left(k_{A B}\right)}} \left[A \ B \ C\right]_{\cal D} = n \epsilon_{A B C} = n \sqrt{\det{\left(\eta_{A B}\right)}} \left[A \ B \ C\right]_{\cal D} \ ,
\eeq
therefore, $f = n^{2/3}$.

\end{document}